# Journal of Geophysical Research: Space Physics

## How the IMF $B_y$ induces a $B_y$ component in the closed magnetosphere and how it leads to asymmetric currents and convection patterns in the two hemispheres

P. Tenfjord[1], N. Østgaard[1], K. Snekvik[1], K. M. Laundal[1], J. P. Reistad[1], S. Haaland[1,2], and S. E. Milan[1,3]

[1]Birkeland Centre for Space Science, Department of Physics and Technology, University of Bergen, Bergen, Norway, [2]Max Planck Institute for Solar System Research, Göttingen, Germany, [3]Department of Physics and Astronomy, University of Leicester, Leicester, UK

**Abstract** We used the Lyon-Fedder-Mobarry global magnetohydrodynamics model to study the effects of the interplanetary magnetic field (IMF) $B_y$ component on the coupling between the solar wind and magnetosphere-ionosphere system. When the IMF reconnects with the terrestrial magnetic field with IMF $B_y \neq 0$, flux transport is asymmetrically distributed between the two hemispheres. We describe how $B_y$ is induced in the closed magnetosphere on both the dayside and nightside and present the governing equations. The magnetosphere imposes asymmetric forces on the ionosphere, and the effects on the ionospheric flow are characterized by distorted convection cell patterns, often referred to as "banana" and "orange" cell patterns. The flux asymmetrically added to the lobes results in a nonuniform induced $B_y$ in the closed magnetosphere. By including the dynamics of the system, we introduce a mechanism that predicts asymmetric Birkeland currents at conjugate foot points. Asymmetric Birkeland currents are created as a consequence of $y$ directed tension contained in the return flow. Associated with these currents, we expect fast localized ionospheric azimuthal flows present in one hemisphere but not necessarily in the other. We also present current density measurements from Active Magnetosphere and Planetary Electrodynamics Response Experiment that are consistent with this picture. We argue that the induced $B_y$ produces asymmetrical Birkeland currents as a consequence of asymmetric stress balance between the hemispheres. Such an asymmetry will also lead to asymmetrical foot points and asymmetries in the azimuthal flow in the ionosphere. These phenomena should therefore be treated in a unified way.

## 1. Introduction

The large-scale dynamics and morphology of the magnetosphere are primarily driven by dayside reconnection between the geomagnetic field and the interplanetary magnetic field (IMF) embedded in the solar wind. *Dungey* [1961] postulated that during southward directed IMF, reconnection would occur on the dayside, merging the terrestrial and the IMF and transferring magnetic flux from the dayside to the nightside. In the magnetotail, open flux transported from the dayside merges again and constitutes a second reconnection region. The cycle is completed when the flux is convected back to the dayside by the return flow. For purely southward directed IMF this mechanism can be assumed to produce symmetric ionospheric convection flows and a near-symmetric configuration in the magnetosphere between the northern and southern lobes.

The solar wind's IMF $B_y$ component is believed to be the cause of a number of asymmetric features in both the magnetosphere and ionosphere [e.g., *Walsh et al.*, 2014]. In the presence of an IMF $B_y$ component the location of the dayside reconnection site on the magnetopause changes (see Figures 1a and 1b or, e.g., *Wing et al.* [1995]). *Park et al.* [2006] found that for a finite IMF $B_y$, the dayside reconnection site moves from the subsolar point, toward high-latitude flanks, and concluded that antiparallel reconnection is dominant over component reconnection for such conditions, illustrated in Figures 1a and 1b. Also, reconnection now produces field lines which no longer convect in a purely antisunward direction but are instead deflected toward the dusk and dawn by the magnetic tension [*Cowley*, 1981]. This additional azimuthal flow on the dayside is added to the antisunward flow produced by the solar wind flow carrying open magnetic field lines.

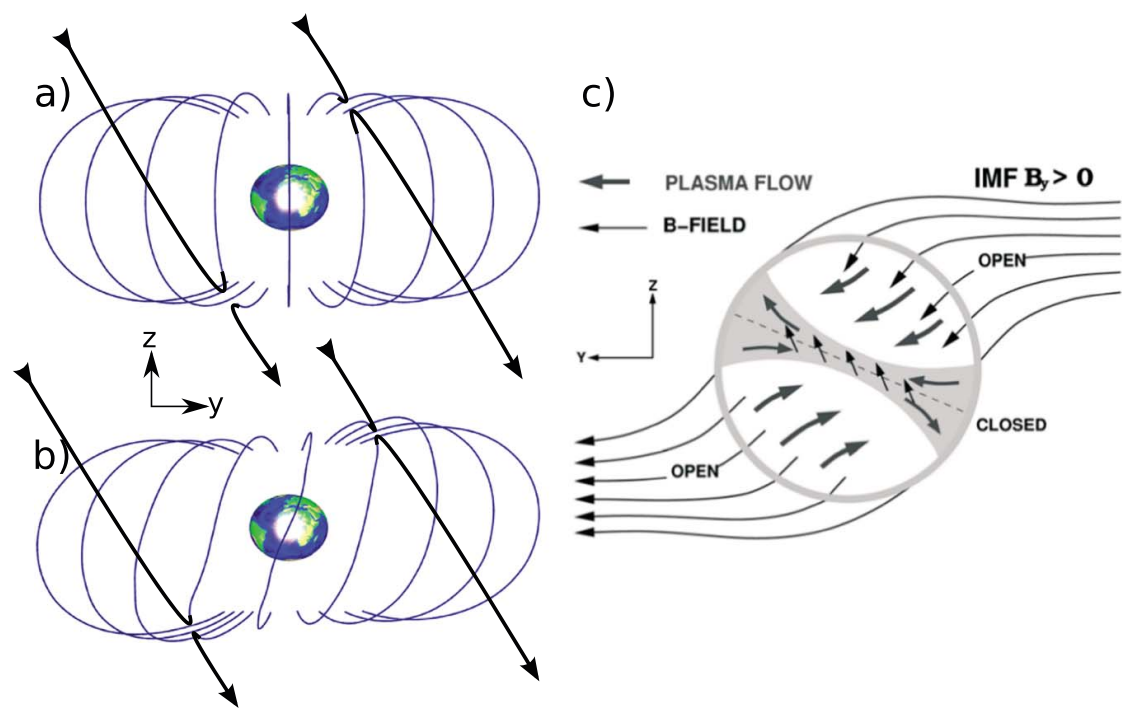

**Figure 1.** (a) Conceptual sketch of IMF (black lines) reconnection with the geomagnetic field (blue lines) when the IMF has a 135° clock angle. The figure serves as an illustration of the topology in the presence of IMF $B_y$. We assume purely antiparallel reconnection [*Park et al.*, 2006]. Reconnection produces field lines which no longer convect in a purely antisunward direction; instead, they are deflected toward the dusk and dawn by the magnetic tension. (b) Reconnection geometry altered by induced $B_y$ on the closed dayside field lines. (c) Asymmetric entry of magnetic flux in the lobes during positive IMF $B_y$ conditions [*Khurana et al.*, 1996; *Liou and Newell*, 2010].

In the ionosphere the original two cells are distorted into "banana"-shaped and "orange"-shaped convection regions. For positive $B_y$ conditions a banana-shaped convection cell is observed in the dawn region and a rounder orange-shaped cell in the dusk region in the Northern Hemisphere. For negative $B_y$ this is reversed. The convection pattern in the Southern Hemisphere is essentially a mirror image of the Northern Hemisphere with respect to the noon-midnight meridian [*Cowley*, 1981; *Provan et al.*, 2009; *Kabin et al.*, 2003]. The influence of IMF sector structure on the magnetic disturbances in the ionosphere was first shown by *Svalgaard* [1968] and *Mansurov* [1969]. Later, *Friis-Christensen et al.* [1972] attributed the observed behavior to the effect of IMF $B_y$.

Inside the closed magnetosphere a $y$ component (or dawn/dusk) of the magnetic field arises, with the same direction as the component in the solar wind. What is often expressed as "penetration" of IMF $B_y$ is a misleading term [e.g., *Kozlovsky*, 2003; *Petrukovich*, 2011]. Instead, we suggest to call it an induced $B_y$ [e.g., *Khurana et al.*, 1996]. The term induced is used to explain the mechanism and processes leading to the $B_y$ component inside the closed magnetosphere. A physical process such as dayside magnetic reconnection is needed to have the external $B_y$ mapped into the system. Through plasma interactions (asymmetric lobe pressure and flux transport) a $B_y$ component of the same sign as the external field is induced in the closed magnetosphere. By calling it induced we imply that it is not simply a result of vacuum superposition, where the magnitude in different regions is determined by the amount of "shielding" from the IMF $B_y$ related penetration electric field, as often seen in the literature [e.g., *Kullen and Blomberg*, 1996; *Walker et al.*, 1999; *Kozlovsky*, 2003]. *Hau and Erickson* [1995] explained the asymmetric velocity $v_y$ in the magnetosphere as due to a north-south electric field component in the solar wind which "penetrates" the tail, thereby causing the flow. The electric field has no power of penetration to drive the motion of the plasma; thus, an externally imposed electric field is unable to "penetrate" into the plasma [*Parker*, 1996; *Song and Vasyliunas*, 2011; *Vasyliunas*, 2012].

How $B_y$ arises in the closed magnetosphere and the consequences are still debated. To our knowledge *Cowley* [1981] was the first to suggest that the post-reconnected field lines on the dayside were added preferentially to different magnetospheric regions (e.g., northern dawn and southern dusk for IMF $B_y > 0$). Whether $B_y$ on closed field lines follows from the reconnection process in the magnetotail [*Hau and Erickson*, 1995; *Cowley*, 1981; *Stenbaek-Nielsen and Otto*, 1997; *Østgaard et al.*, 2004] or if the asymmetric loading of open flux also influences closed field lines [*Khurana et al.*, 1996] is still debated. *Khurana et al.* [1996] suggested that through asymmetric loading of flux into the different magnetospheric lobes, the shear flow ($y$ directed) between northern and southern halves of the plasma sheet could generate a $B_y$ component on closed field lines (see Figure 1c).

A relationship between interplanetary magnetic field (IMF) $B_y$ and an induced $B_y$ component in different magnetospheric regions has been established statistically [*Fairfield*, 1979; *Cowley and Hughes*, 1983; *Lui*, 1984; *Kaymaz et al.*, 1994; *Cao et al.*, 2014; *Stenbaek-Nielsen and Otto*, 1997; *Wing et al.*, 1995; *Petrukovich et al.*, 2005] and shows that the induced $B_y$ is not distributed uniformly in the closed magnetosphere. *Wing et al.* [1995] found that a fraction of the IMF $B_y$ component appeared at all local times but stressed that it was strongest near local noon and midnight. *Kaymaz et al.* [1994] showed that the induced $B_y$ can be as large as 35% of the IMF $B_y$ at the flanks, compared to 26% at the central portion of the plasma sheet. *Stenbaek-Nielsen and Otto* [1997] argued that during the evolution of a flux tube moving from the tail toward the Earth, flux will be accumulated (pileup) and thereby generate a region of enhanced $B_y$. We note that the asymmetries exist on both open and closed field lines.

The effects of the induced $B_y$ have been extensively studied. The substorm onset location [*Liou and Newell*, 2010; *Liou et al.*, 2001; *Østgaard et al.*, 2011a] has been found to exhibit a longitudinal dependence on the presence of IMF $B_y$. A number of auroral studies have shown that there are systematic displacements and intensity differences [e.g., *Reistad et al.*, 2014; *Cowley*, 1981] in the aurora in the two hemispheres. The longitudinal displacement of aurora between the two hemispheres has been shown to correlate with IMF $B_y$ [see *Østgaard et al.*, 2007, and references therein]. It is now generally accepted that aurora is a manifestation of Birkeland (field-aligned) currents [*Strangeway*, 2012]. Based on concurrent observations of the IMF orientation, *Stenbaek-Nielsen and Otto* [1997] have proposed a mechanism for how IMF $B_y$ can give rise to interhemispheric currents between the two hemispheres. We present and discuss their mechanism in section 3. *Østgaard and Laundal* [2012] proposed that this mechanism could explain some of the nonconjugate auroral observations.

*Milan* [2015] has previously discussed the contributions of magnetic tension forces and asymmetrical loading of the lobes with new open flux to produce dawn-dusk asymmetries in flows in the magnetosphere-ionosphere system; the current paper investigates these influences from a mathematical and modeling perspective. Our motivation for the present work is to address how the presence of IMF $B_y$ changes the dynamics and configuration in the magnetosphere. Furthermore, we investigate the consequences that follow in terms of induced $B_y$, asymmetric Birkeland currents, and associated convection patterns. We believe that these asymmetric azimuthal flows in both the dayside and nightside should be accompanied with nonconjugate aurora.

The paper is organized as follows: section 2 gives a theoretical background of the forces responsible for the evolution of $B_y$ and $v_y$ and their interdependency. Section 3 describes how IMF $B_y$ induces $B_y$ on closed field lines in the magnetosphere. We show how the forces are distributed and describe which forces are dominating in the different regions. We also discuss how convection cell patterns are modulated by the presence of IMF $B_y$. Results from our magnetohydrodynamics (MHD) model run are related to the equations introduced in section 2. In section 4 we introduce a mechanism that predicts asymmetric Birkeland currents on the same field line and argue that these are pairs of Birkeland currents instead of "interhemispheric" currents. In section 5 we present statistical data from Active Magnetosphere and Planetary Electrodynamics Response Experiment (AMPERE) and compare it to expected signatures from our mechanism. In section 6 we discuss the response time of the nightside magnetosphere with respect to the arrival of IMF $B_y$ at the dayside magnetopause.

## 2. Theoretical Background

In this section we describe the governing MHD equations relevant for understanding how $B_y$ is induced in the magnetosphere.

MHD can be expressed by a set of coupled interdependent dynamical equations. We use the momentum equation to determine how the forces can affect the flow and show how asymmetric transport of flux leads to an induced $B_y$. This induced $B_y$ can again of course affect the flow, as the momentum equation is coupled with the induction equation. Apart from energy dissipation through Ohmic heating, magnetic stress remains stored in the system in the form of magnetic energy. The induced $B_y$ in the magnetosphere can therefore be explained as a simple shoving match between stresses. The momentum equation can be written as [cf. *Parker*, 1996, 2007]

$$\rho \frac{dv_i}{dt} = \frac{\partial}{\partial x_j} M_{ij} - \frac{\partial}{\partial x_j} P_{ij}$$

$$= \frac{1}{\mu_0} \left( \frac{\partial}{\partial x_j} (B_i B_j) - \frac{1}{2} \delta_{ij} \frac{\partial}{\partial x_j} B^2 \right) - \frac{\partial}{\partial x_j} P_{ij} \tag{1}$$

where $M_{ij}$, $P_{ij}$, $\rho$, and $v$ are the Maxwell stress tensor, pressure tensor, mass density, and velocity, respectively. The induced $B_y$ in the magnetosphere can be understood in terms of the magnetic forces acting on plasma. In the following equations we consider how the magnetic field and the flow are modulated in the $y$ direction. We focus on the $y$ components since the additional forcing arising from IMF $B_y$ acts primarily in this direction. The evolution of $v_y$, considering electromagnetic forces alone, is (assuming that the plasma pressure in the lobes is negligible)

$$\rho \frac{dv_y}{dt} = \frac{1}{\mu_0} \left( B_x \frac{\partial}{\partial x} + B_z \frac{\partial}{\partial z} \right) B_y - \frac{1}{2\mu_0} \frac{\partial}{\partial y} \left( B_x^2 + B_z^2 \right) \tag{2}$$

$M_{ij}$ has been expanded to four terms, the two first terms describe the tension along the field lines, and the two last magnetic pressure. The tension is related to the field line curvature: $\vec{T} = \frac{\partial}{\partial x_j}(B_i B_j) = B^2 \frac{\hat{n}}{R_C}$, where $R_C$ is the radius of curvature of the field line and $\hat{n}$ is unit vector pointing away from the center of curvature. For self-consistency we also need an equation to determine the evolution of the magnetic field, found by combining the Maxwell-Faraday equation with the conservation of mass:

$$\frac{d\vec{B}}{dt} = (\vec{B} \cdot \nabla)\vec{v} + \frac{\vec{B}}{\rho} \frac{d\rho}{dt} \tag{3}$$

where we have used $\vec{E} = -\vec{v} \times \vec{B}$ and the convective derivative is defined as $\frac{d}{dt} = \frac{\partial}{\partial t} + \vec{v} \cdot \nabla$. The $y$ component of equation (3) can be expressed as

$$\frac{dB_y}{dt} = \left( B_x \frac{\partial}{\partial x} + B_y \frac{\partial}{\partial y} + B_z \frac{\partial}{\partial z} \right) v_y + \frac{B_y}{\rho} \frac{d\rho}{dt} \tag{4}$$

The three first terms on the right-hand side describe how $v_y$ can change the magnetic field over time. These terms dominate when the assumption of incompressible flow holds. The assumption should be valid as long as density gradients are small (uniform flow) and the driving of the system can be considered steady (steady flow). We consider these terms to dominate in the outer magnetosphere [*Escoubet et al.*, 1997; *Laakso et al.*, 2002]. Considering regions where $B_y$ has not yet been induced, the second term is negligible. In these regions, the induced $B_y$ is attributed to the three remaining terms. The shear flow (flow induced by a force), $v_y$, is determined by equation (1).

The fourth term tells us that the change of $B$ can be caused by compressing or expanding the plasma [*Hau and Erickson*, 1995]. This term becomes important in the inner magnetosphere. This can be used to explain what is called "pileup" of magnetic flux. As a magnetic field line is convected closer to Earth from the tail, the mass density increases with the magnetic field strength. *Stenbaek-Nielsen and Otto* [1997] suggested that since the magnetic field strength increases as the field lines convect earthward, and consequently also its $B_y$ component, an interhemispheric current would arise due to the gradient in the $x$ direction of the $B_y$. However, we will argue that this is not correct.

## 3. Generation of $B_y$ in the Magnetosphere

In this section we present results from the Lyon-Fedder-Mobarry (LFM) [*Lyon et al.*, 2004; *Merkin and Lyon*, 2010] global MHD model. We interpret the results in terms of the forces acting to produce a $B_y$ component in the magnetosphere and relate them to the equations introduced in section 2. The LFM model provides a self-consistent model of the global magnetosphere. Even though the model is based on the equations of ideal MHD, it has proven to be an extremely useful tool for studying the large-scale dynamics of the magnetospheric system [*Ridley et al.*, 2010]. Due to the lack of resistivity, magnetic reconnection does not exist in ideal MHD. Instead, diffusion is introduced by numerical effects, which comes about when magnetic field gradient scale length approaches the computational grid size [*Lyon et al.*, 2004; *Ouellette et al.*, 2010]. The model has several limitations, especially in the inner magnetosphere and in the self-consistency of the magnetosphere-ionosphere coupling [see *Ridley et al.*, 2010; *Tu et al.*, 2014].

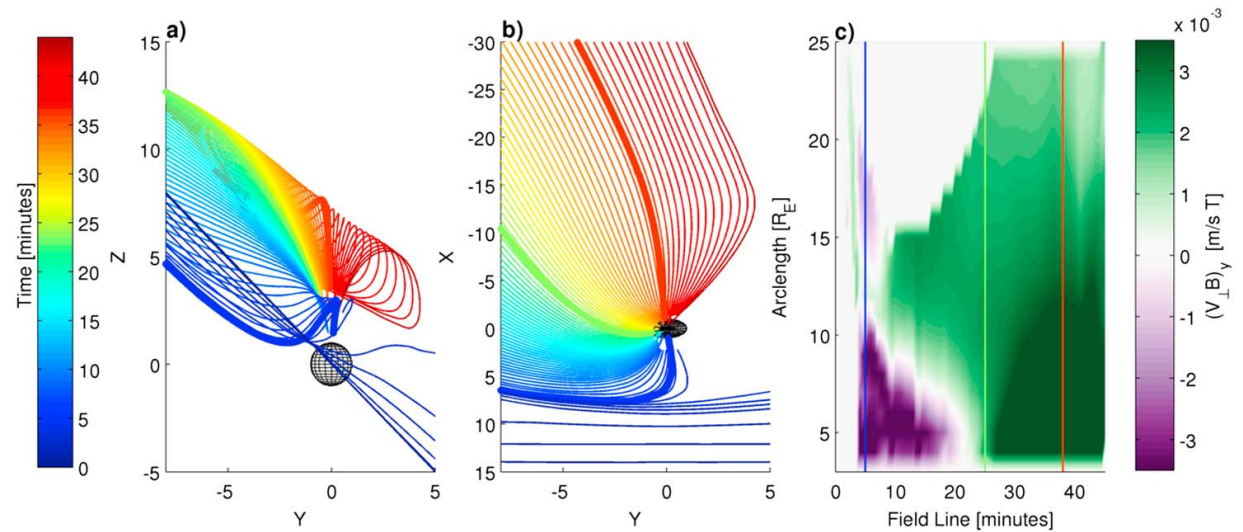

**Figure 2.** The evolution of a field line from the solar wind to the nightside reconnection region, (a) seen from the Sun and (b) down on the Northern Hemisphere. The color bar represents time with 30 s between each field line. Immediately after dayside reconnection, the field lines are forced dawnward by the tension (equation (2) and Figures 1a and 1b). After about 20 min the field is forced duskward by the accumulated pressure in the northern dawn lobe (Figure 1c). (c) The flux transport in the $y$ direction, $(v_\perp B_y)$. The three vertical lines are colored in accord with the time code: blue, green, and red (shown as bold colored lines in Figures 1a and 1b) represent positions during the trace, for visualizing purpose only. The first 20 min the flux transport is directed dawnward, corresponding to the tension, followed by duskward transport due to pressure.

The solar wind conditions during the model run are $B_Z = -10$ nT and $V_X = -400$ km/s with zero dipole tilt. The IMF $B_y$ is zero on the first 30 min of the run, followed by $B_y = 10$ nT on the remaining 2.5 h. Start time ($t = 0$) is defined as the time IMF $B_y$ arrives at the dayside magnetopause. The data have been produced by Community Coordinated Modeling Center (CCMC) and are available as run number Paul-Tenfjord-032514-1.

We trace field lines as they evolve in time by using their perpendicular velocity at some initial time to calculate the location of the field line 30 s later. We also project the foot points of the field lines, by tracing the field lines to 3 $R_E$ (model constraint) radius and use simple dipole mapping from this altitude. We acknowledge that the method relies on the field not departing from a dipole configuration between the surface and the 3 $R_E$ altitude. We use the method to quantify the asymmetry of the foot points, defined to be the deviation (in $\Delta$MLT) from a purely symmetric configuration. The deviation can be both latitudinal and longitudinal; we focus on the latter, often associated with IMF $B_y$. We discuss the dayside and the nightside separately.

### 3.1. $B_y$ on the Dayside
In Figures 2a and 2b we follow a field line in the Northern Hemisphere from the solar wind to the nightside as it evolves over time. The colors indicate time, with 30 s between each field line. Figure 2a shows the field lines in the $Y$-$Z$ plane. The solar wind approaches the magnetopause with a clock angle of 135°.

After dayside reconnection the solar wind plasma carries open magnetic field lines with it, while the foot points of these field lines remain anchored to the Earth. The transverse momentum from the solar wind flow is transmitted to the foot points such that they eventually move laterally.

Due to IMF $B_y$, there exists an additional tension in the field line just after the dayside reconnection. The tension (related to the field line curvature: $\vec{T} = \frac{\partial}{\partial x_j}(B_i B_j) = B^2 \frac{\hat{n}}{R_C}$; see equation (2)) is directed dawnward in the Northern Hemisphere ($\frac{dv_y}{dt} < 0$) and duskward in the Southern Hemisphere as shown in Figures 1a and 1b. Figure 2b shows the view in the $Y$-$X$ plane. Due to the reconnection geometry of the terrestrial magnetic field, the tension is stronger on the dusk side compared to the dawn side in the Northern Hemisphere (Figures 1a and 1b). The tension term of equation (2) dominates over the pressure term. Immediately after dayside reconnection, the tension along $x$ changes $v_y$ as described by the momentum equation (for Northern Hemisphere: $B_x \frac{\partial B_y}{\partial x} < 0$, first term in equation (2)).

Figure 2c shows flux transport in the $y$ direction along the arc length of each of the field lines. The tension acts on the newly reconnected open field lines on the dayside at about ∼10 $R_E$ (arc length) during the first 8 min, seen in purple in Figure 2c. The part of the field lines outside the magnetosphere is not affected (top left). This purple region extends earthward; this is the tension propagating along the field line earthward accelerating the plasma in the dawn direction. This Maxwell stress exerted on the ionosphere eventually moves the ionospheric foot points in the dawnward direction, seen as purple between 5 and 20 min.

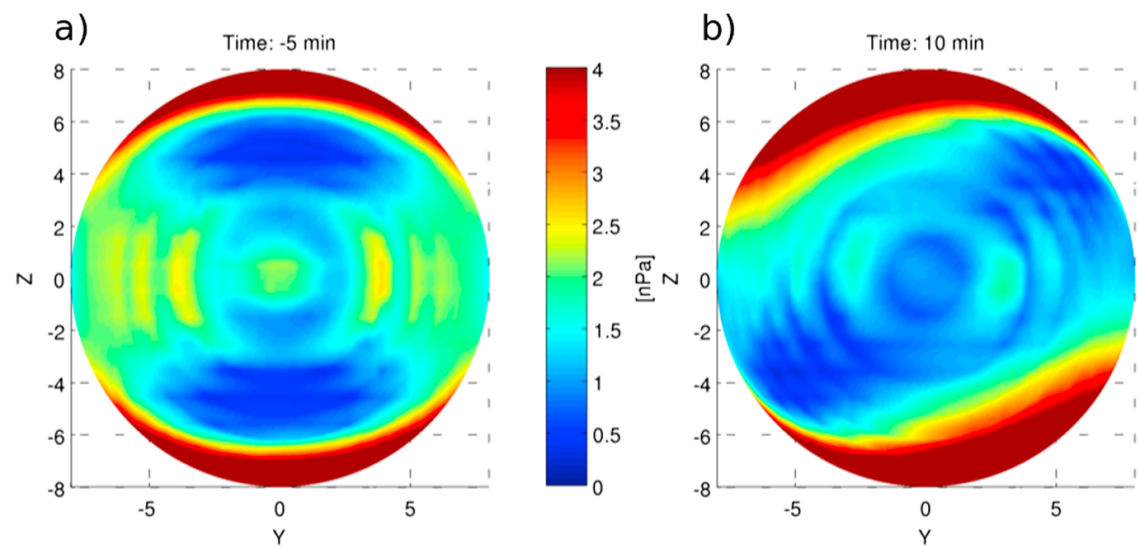

**Figure 3.** Magnetic pressure projected on a half sphere of size $R = 8\,R_E$, so that the center is at $X = 8\,R_E$, but the outmost values are at $X = 0\,R_E$. (a) Magnetic pressure 5 min prior to IMF $B_y = 10$ nT. (b) Ten minutes after IMF $B_y$ arrived at magnetopause.

After about 10 min, the magnetospheric part of the field line (arc length between 10 and 15 $R_E$) starts to experience the enhanced pressure in the northern lobe. The magnetic pressure in the lobes is a result of asymmetric loading of flux. The region of enhanced pressure is confined by the constant external stresses applied by the magnetosheath flow against the magnetopause, essentially maintaining a circular cross section of the magnetotail. This localized region of enhanced pressure (seen in green at about 10 min in Figure 2c) will start to force the magnetospheric plasma in the dusk direction $((v_\perp B)_y > 0)$. This transverse momentum is transmitted to the ionosphere, forcing the ionospheric foot points in the dusk direction after about 25 min. The field lines are now forced duskward and at the same time converge toward the neutral sheet (see Figure 1c). Eventually, they will reconnect with the approaching field lines from the southern lobes. The vertical colored lines in Figure 2c are added for visualization purpose only, and they represent the three field lines with larger line width in Figures 2a and 2b.

Figure 3 shows the magnetic pressure presented on a half sphere with radius 8 $R_E$ looking from the Sun toward Earth. Figure 3a shows that the pressure distribution 5 min prior to IMF $B_y$ has been introduced to the MHD model. Figure 3b shows that the configuration 10 min after IMF $B_y = 10$ nT. Figure 3b suggests that more flux is eroded from the northern dusk and southern dawn high-latitude regions and added asymmetrically to the northern dawn and southern dusk, respectively. The enhanced pressure in the northern dawn and southern dusk will also displace the existing closed field lines in the region. In the northern dawn and southern dusk the closed field lines will be compressed equatorward. The northern dusk and southern dawn are not affected. This asymmetric forcing will induce a latitudinal asymmetry in the foot points, as already noted in *Cowley et al.* [1991] (see their Figure 2). Also associated with these asymmetric forces are the twisting of the dayside field lines, which is analogous to an induced $B_y$. We agree with the conclusion of *Wing et al.* [1995] that the erosion of magnetic flux at the high-latitude flanks on the dayside, combined with the newly opened flux added to the dawn (for Northern Hemisphere, see Figure 3), will induce a $B_y$ on the dayside. This in turn affects the dayside reconnection by changing the reconnection geometry. The altered reconnection geometry is shown in Figure 1b. The IMF reconnects with a twisted terrestrial field, causing an even greater tension on the newly reconnected field lines; see Figure 1b.

To summarize the effect of a positive IMF $B_y$ component on the dayside, the field lines are forced both dawnward (by the tension) and at the same time tailward by the solar wind. Considerably more flux is added to the dawn side of the noon-midnight meridian compared to the dusk side in the Northern Hemisphere. The closed field lines on the dayside get twisted by the combination of erosion of flux at high-latitude flanks and by the increase of magnetic pressure in the northern dawn and southern dusk. The induced $B_y$ on the dayside has the same sign as the IMF as sketched in Figure 1b and shown by *Burch et al.* [1985].

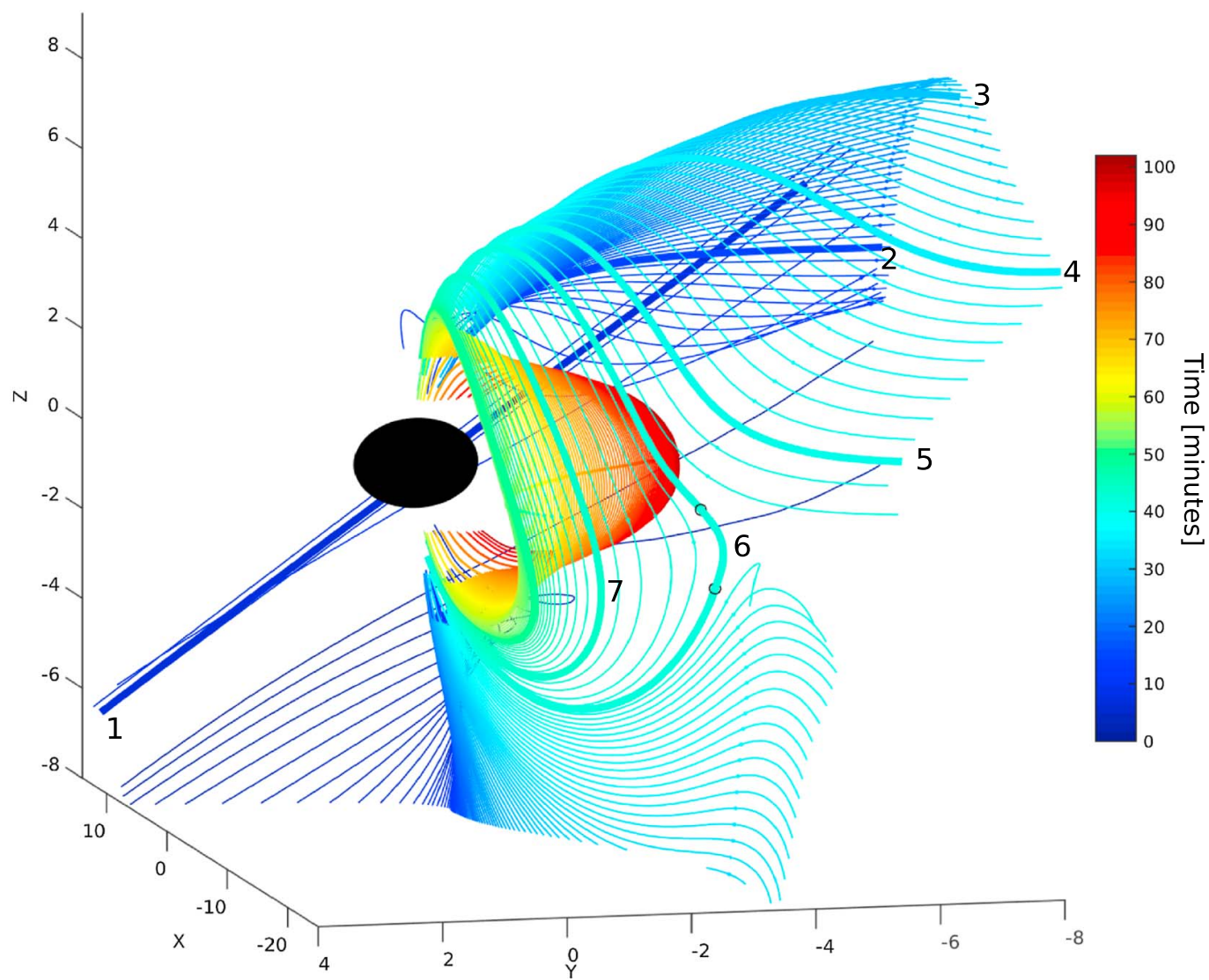

**Figure 4.** Illustration of a field line as it evolves over time from dayside reconnection (1), convection in the lobe (2–5), nightside reconnection (6), and earthward convection (7) (return flow) on dawn cell. Our starting point of the trace is the field line at (6); from this approximate reconnection site we trace backward in time (1–5) and forward in (6–7). Colors represent time. It takes approximately 45 min from dayside reconnection until the field reconnects again in the magnetotail. From the start of the return flow, we follow the field lines about 50 min until the foot points become symmetric again.

### 3.2. Nightside $B_y$

After this first phase, approximately after the field line has crossed the terminator plane, the magnetic pressure terms in equation (2) dominate (see Figure 2). In this second phase, the accumulated flux in the northern dawn and southern dusk dominates the evolution of $v_y$. For simplicity we can assume that all the $y$ directed stress after the dayside reconnection is stored in the system as magnetic pressure (under the assumptions that the tension is removed in the passage from the dayside to the lobe, the magnetospheric pressure is confined and Joule heating is negligible).

The region of enhanced magnetic flux in the northern dawn and southern dusk is localized close to the terminator plane. The magnetic pressure localized in this region immediately forces both the incoming field lines and the surrounding field lines to move. The accelerated plasma will extend spatially as a direct result of the magnetosphere's attempt to restore pressure balance. The distribution of induced $B_y$ in the magnetotail (and on the dayside) is a consequence of magnetic pressure accelerating the plasma. Even though it may appear that the asymmetric magnetic pressure is distributed far downtail, it is, in fact, the accelerated plasma that extends, inducing $B_y$ as the plasma propagates. The magnetic field is transported with the plasma in the dusk (north) and dawn directions (south) in the magnetotail for IMF $B_y > 0$. At the same time the flux tubes are convected toward the neutral sheet. This means that the plasma propagates toward the neutral sheet at some angle (see Figures 4 and 1c). As this compression extends, it also affects the surrounding closed field lines, thereby inducing a $B_y$ in the tail, independently of a reconnection process. This has previously been suggested by *Khurana et al.* [1996], as shown in Figure 1c. By the same arguments, this is also how the dayside induced $B_y$ arises. We also note that since the dayside field lines are twisted, the increased tension of the newly reconnected field lines results in an even greater magnetic pressure added to the lobes (see Figure 1b).

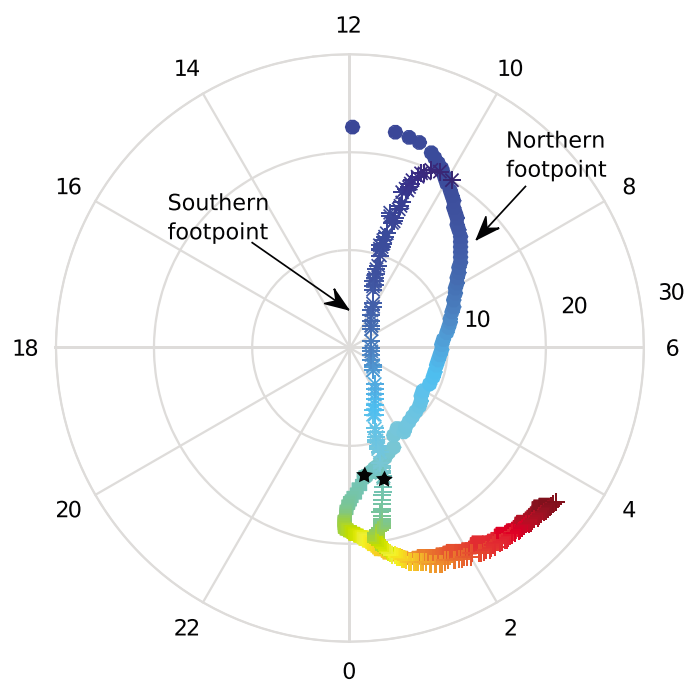

**Figure 5.** Corresponding foot points for Figure 5. Filled circles represent foot points in Northern Hemisphere and asterisk (∗) for Southern Hemisphere. The two black stars mark the time step when field lines go from open to close. The colors follow the colors in Figure 4. The flow (evolution of the foot point) in the north coincides with the crescent-shaped banana convection cell and is forced further toward dusk compared to the southern foot point. Eventually, as the foot points are forced toward lower latitudes (magnetospheric flow earthward), the northern foot points catch up with the southern foot points and they continue their convection toward the dawn-dusk meridian with symmetric foot points.

In Figure 4 we show the evolution of a newly reconnected field line (6) in the magnetotail viewed from the tail and rotated slightly toward dusk. From this approximate reconnection point we trace backward and forward in time. Following the field lines backward in time (1–6), we trace them to the approximate same field line in the solar wind. However, we note that the convection over the Southern and Northern Hemispheres does not take the exact same amount of time, and we can therefore not claim that it is the same field line. Tracing forward in time (6–7), we observe an earthward and dawnward convection.

In Figure 5 we present the projected foot points of the field lines in Figure 4. An asterisk (∗) represents the foot points in Southern Hemisphere and filled circles the Northern Hemisphere. As mentioned the field lines start off at the approximate same field line in the solar wind. The foot points on the dayside after reconnection in Northern and Southern Hemispheres have asymmetric positions of ∼2 h (12 h and 10 h magnetic local time (MLT), respectively), which is a signature of the twisted dayside magnetic field. Immediately after dayside reconnection, the foot point in the Northern Hemisphere convects dawnward, while the foot point in the Southern Hemisphere moves duskward. After about 25 min (see Figure 2c) the northern foot point moves in the duskward direction; directions are reversed in the Southern Hemisphere. As the field lines approach the neutral sheet (4–5 in Figure 4), at the angle dictated by the pressure distribution, the field lines reconnect. In Figure 5 the two black markers (∼78° latitude) show the position of the start of the trace, which is a newly reconnected field line (6 in Figure 4).

They reconnect with asymmetric foot points. As they dipolarize and convect earthward (6–7 in Figure 4), the asymmetry increases, which is due to the distribution of the pressure forces. Eventually, the flux tubes are forced either dawnward or duskward by the plasma and magnetic pressure forces surrounding the Earth. As the flux tube starts to convect dawnward (green color markers), the foot point in the Northern Hemisphere is around 0 MLT, while the southern foot point is at 1 MLT. As time evolves, the foot points become more symmetric again, meaning that the azimuthal flow in the north has caught up with azimuthal flow in the Southern Hemisphere. Asymmetric azimuthal flows represent the large-scale rectifying of a twisted flux tube, which necessarily need to be accompanied by asymmetric Birkeland currents. In the next section, we will discuss how these differences in azimuthal flow relate to Birkeland currents on conjugate field lines.

Before we summarize the effects of IMF $B_y$ on the nightside, the following question arises: how does reconnection manipulate the system, and which field lines actually reconnect? As shown above it is the force balance in the magnetosphere that dictates where the field lines from the different lobes converge. That is, the pressure distribution in the tail determines the direction in which the flow approaches the neutral sheet. As shown in Figure 5 the global MHD model suggests that they do indeed reconnect with asymmetric foot points. We acknowledge the uncertainty in both the location, tracing and of the model, in general, regarding details. There are uncertainties in the diagnostic methods which are very difficult to quantify; for instance, small errors may link two different field lines in the tracing procedure [*Song and Vasyliūnas*, 2010]. The errors could also accumulate along the trace resulting in a large uncertainty. This is especially important in reconnection regions where the topology rapidly changes. Even though not able to quantify the uncertainty, we are confident that foot points in Figure 5 describe the behavior properly for the following reasons: (1) the evolution of the foot points is smooth and consistent. (2) Retracing the field lines after a small change in the location and a temporal shift reveals a comparable pattern. Nevertheless, our main point is the role of the pressure distribution, forcing the flux tubes in the y direction asymmetrically between the lobes. We interpret the morphology of reconnected field lines as a consequence of these asymmetric forces. We emphasize that the newly reconnected flux tube will continue to experience these forces, and in addition, the magnetic field strength increases as the field line dipolarize according to equation (4). This means that closed field lines present before IMF $B_y$ arrived will also have an induced $B_y$ independently of tail reconnection. In the next section we discuss this more thoroughly, including the resulting Birkeland currents.

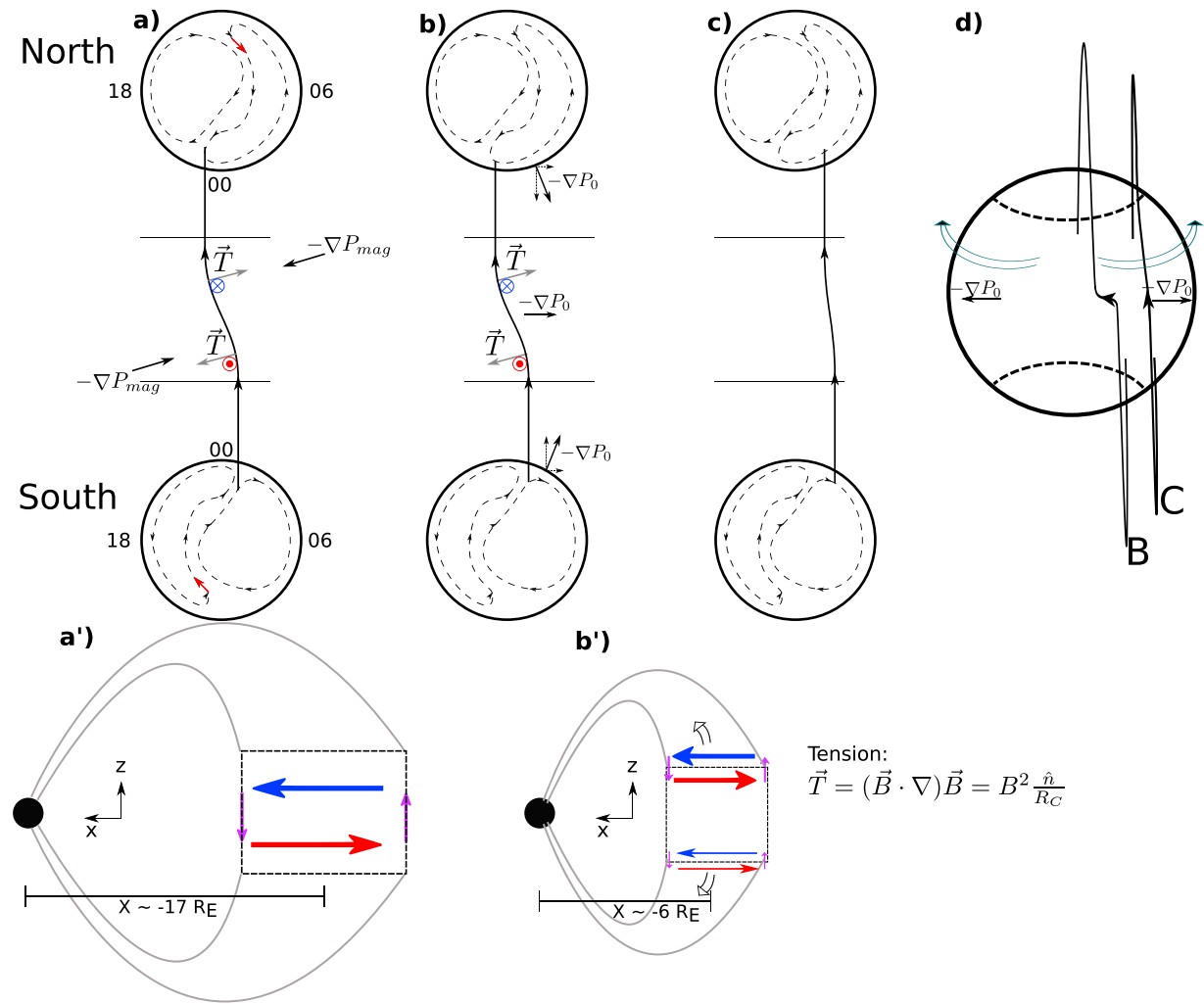

**Figure 6.** A flux tube on closed field lines with asymmetric foot points convecting on the dawn convection cell during IMF $B_y > 0$ conditions. (a–c) Pressure, tension, and asymmetric foot points into the dawn cell. (a′ and b′) The associated current systems seen from dusk. Figure 6a shows closed field line in midtail region around $X \sim -17\ R_E$ experiencing the asymmetric pressure from the lobes $\nabla P_{mag}$ which is balanced by the tension $\vec{T}$. Figure 6a′ shows currents closed locally. Figure 6b shows that field line convects earthward and is affected by the pressure (plasma and magnetic) surrounding Earth $\nabla P_0$. Now the forces do not balance. In the Northern Hemisphere these forces point in the same direction. Hence, most of the stress is transmitted into this hemisphere, and the northern foot point will catch up with the southern counterpart to restore symmetry. Figure 6b′ shows a stronger current system deposited into the Northern Hemisphere, compared to the Southern Hemisphere. Figure 6c shows the Northern Hemisphere catching up with the southern foot point, making them symmetric. (d) View from magnetotail toward the Sun and shows the relaxation of the field line from Figures 6b to 6c.

### 3.3. Ionospheric Convection

In this section we review the consequences of the above asymmetric magnetospheric forcing for the properties of flows in the ionosphere. An illustration of the convection patterns for Northern and Southern Hemispheres is shown in Figure 6a for IMF $B_y > 0$. In the presence of IMF $B_y$, the normal two-cell convection pattern is modified into what is known as banana and orange cells [*Heppner*, 1972; *Mozer et al.*, 1974; *Heppner and Maynard*, 1987; *Cowley et al.*, 1991; *Haaland et al.*, 2007; *Grocott and Milan*, 2014]. A crescent banana convection cell is seen on the dawn cell in the Northern Hemisphere, while the convection cell on the dusk side has a more rounded orange shape (opposite for Southern Hemisphere). The banana-shaped convection cell can extend into the premidnight and postnoon regions (see Figure 6a and Figure 5).

The shape of convection cells is determined by the imposed force from the magnetosphere. On the dayside the tension on the newly opened field lines is forced dawnward in the Northern Hemisphere and southward in the Southern Hemisphere; see red arrows in Figure 6a. The associated $\vec{J} \times \vec{B}$ force is transmitted down to the ionosphere accelerating the ionospheric plasma in the direction dictated by the imposed force, resulting in the crescent dawn cell and rounded dusk cell in the Northern Hemisphere. On the nightside it is the pressure distribution in the lobes that forces the ionospheric convection duskward. When we follow a field line from the nightside reconnection region as it convects toward Earth, the pressure ($\nabla P_0$-plasma plus magnetic) surrounding Earth will force the field line to convect either on the dawn cell or the dusk cell. In the next section we discuss the associated current system required when the flow proceeds on a banana cell in the Northern Hemisphere and on an orange cell in the Southern Hemisphere. This corresponds to the dawn cell for IMF $B_y > 0$.

Finally, we notice that we have considered dayside reconnection alone as the source of the asymmetric pressure distribution in the magnetosphere, which in turn induces $B_y$. In fact, there are other mechanisms that can produce $B_y$, such as warping of the plasma sheet [*Russel*, 1972; *Tsyganenko*, 1998; *Liou and Newell*, 2010], lobe reconnection, magnetotail twisting, and tilt effects [*Petrukovich et al.*, 2005; *Petrukovich*, 2011]. Also, we have not included nor discussed any consequences of ionospheric properties such as conductivity, conductivity gradients, and/or other ion-neutral interactions. We note that the effects of these processes may amplify or mitigate the signatures of the mechanisms.

## 4. Generation of Birkeland Currents Due To $B_y$

In this section we discuss the asymmetric Birkeland currents on the nightside, arising as a consequence of an induced $B_y$ component in the magnetosphere. By definition, Birkeland currents transmit the transverse momentum (i.e., magnetic tangential stress) and energy from the source region to the region of dissipation, along magnetic field lines [*Iijima*, 2000].

That is, the Birkeland currents are a consequence of perpendicular (to $B$) perturbations along the magnetic field lines. The closure current (Pedersen current) is at the wavefront of the associated Alfvèn wave. The electric current is a result of the interplay of the forces in the momentum equation. Since the associated electric fields and current play no role in the dynamics (they are created and driven by the varying $B$ and $v$ [*Parker*, 2007]), we argue that the ionospheric signatures are better understood in terms of forces and flows. For this reason it is reasonable to discuss the force balance in context of the Birkeland currents. The Birkeland currents transport the tangential stress from the source in the tail, which is a result of asymmetric loading of flux.

### 4.1. Asymmetric Birkeland Currents

We now present a framework in which we predict asymmetric Birkeland currents as a result of induced $B_y$.

In Figures 6a–6c we show a conceptual illustration of a field line in the magnetotail at different stages of its evolution. We consider a field line that convects along the dawn cell.

In this illustration $B_y$ is defined as finite and positive inside the boxes (Figures 6a′ and 6b′) and zero elsewhere. The gradients of $B_y$ along the $x$ direction are clearly defined as step functions at the boundaries, and the resulting Birkeland currents (violet arrows) are represented as infinitesimally thin sheets (Figures 6a′ and 6b′). In reality the limits of the perturbation are not so impulse like and distributed over a larger area.

Figures 6a and 6a′ show a closed, midtail field line in the magnetotail region. The asymmetric loading of flux to the lobes exerts an asymmetric pressure on the field line, directed duskward in the northern lobe, and dawnward in the southern lobe. The foot points of the field line are asymmetric. The field line is twisted by the magnetic pressure. An opposing tension acts to balance the magnetic pressure $-\nabla P_{\text{mag}}$ (Figure 6a and equation (2)). These forces could, in general, balance, resulting in a force-free configuration. The current system of such an equilibrium is shown in Figure 6a′. In this situation, the current system is locally closed, the perturbation is not propagating, and currents are dissipationless. In the midtail region, the closed field lines are highly nondipolar; magnetic tension forces the field lines earthward; see Figures 6b and 6b′. Due to the finite extent of the magnetic pressure distribution in the lobes, the force acting on the flux tube is now dominated by the combined magnetic and particle pressure surrounding Earth, $-\nabla P_0$. This force acts radially outward from Earth. The twisted field line can now start to relax. In Figure 6b, $-\nabla P_0$ is directed parallel to the tension in the northern lobes and antiparallel to the tension in the southern lobe. The current system for this situation is shown in Figure 6b′. The northern part of the field line deposits the majority of the tension stored in the flux tube to the Northern Hemisphere, since $-\nabla P_0$ and the tension $\vec{T}$ act in the same direction (Figure 6b). The current system propagating down to the northern ionosphere is therefore larger than in the Southern Hemisphere. We stress that the current systems illustrated here are isolated from the "normal-" driven Region 1 (R1) and Region 2 (R2) systems. If the current system belonging to Figure 6b were to be superimposed with the return flow-generated R1 and R2 currents, the result would be two current systems with equal (symmetric) direction but with different magnitude. Figure 6c shows the situation when the flux tube has relaxed and the foot points are again symmetric. In this configuration the foot point in the Northern Hemisphere has moved faster to catch up with the foot point in the south. That is because the magnetosphere imposed a stronger flow downward in the northern lobe (as we showed in Figure 5), resulting in a stronger Pedersen current to propagate down which in turn accelerates the ionospheric plasma. The flux tube would then continue to convect without tension and with symmetric foot points toward the dayside. Figure 6d shows the flux tube as viewed from the tail toward the Sun, corresponding to Figures 6b and 6c. Here we have emphasized that the northern foot point will move a larger distance and hence faster, compared to the Southern Hemisphere.

It is important to include the dynamics of the mechanism. During the evolution of the twisted field line it experiences different forces (varying amount of pressure gradient and curvature and Earth's surrounding pressure), and only when one views these different interactions in a unified way are we able to describe the mechanism properly.

Consequently, one would expect to observe a pair of asymmetric Birkeland currents connecting to the two hemispheres, not necessarily crossing the neutral sheet. The source of the tangential stress is in the magnetotail region, and it is oppositely directed in the northern and southern lobes. The stress from the source region is transmitted to the ionosphere via Alfvèn waves. The force balance exists between each ionosphere and the magnetotail separately. The forces and energy may be distributed asymmetrically; for IMF $B_y > 0$ a larger stress will be transmitted to the northern postmidnight region, and southern premidnight, which should result in asymmetries in both currents, azimuthal flows, and auroras.

We note that each ionosphere may respond differently to applied magnetospheric stress owing to differences in ionospheric properties [*Tu et al.*, 2014].

### 4.2. Interhemispheric Currents

In this section we present the proposed mechanism by *Stenbaek-Nielsen and Otto* [1997], where they suggest that observed hemispherical difference in the aurora is a consequence of interhemispheric currents. An interhemispheric current is defined as a current flowing between two conjugate ionospheres [*Lyatskaya et al.*, 2014]. We also discuss why we do not agree with their conclusions on the existence of an interhemispheric current driven by the magnetosphere.

As first suggested by *Stenbaek-Nielsen and Otto* [1997], the induced $B_y$ is not distributed uniformly in the closed magnetosphere but tends to increase toward the Earth due to the pileup of magnetic flux toward the inner edge of the plasma sheet, as illustrated in Figure 7. This has been shown empirically. On the nightside, *Wing et al.* [1995] found that 79% of the IMF $B_y$ induces a $B_y$ component at $X = -6.6\ R_E$, for $-30 < X < -10\ R_E$ *Lui* [1984] found 50%, and further downtail between $-33 < X < -20\ R_E$ *Fairfield* [1979] found the induced $B_y$ component to be 13% of the IMF $B_y$ strength. This can be understood in terms of the forth term in equation (4); for steady flow and no shear flows the change of $B_y$ is given by $\frac{dB_y}{dt} = \vec{v} \cdot \nabla \rho$. The second assumption would be valid if $B_y$ on closed field lines on the nightside was due to reconnection with asymmetric foot points only [e.g., *Stenbaek-Nielsen and Otto*, 1997]. If so, as the closed field lines convect earthward, they would experience a positive gradient in the density and thereby intensify $B_y$ (along with $B_x$ and $B_z$). However, we argue that $B_y$ is induced by the shear motion created by the asymmetric loading of flux from the dayside, and therefore, the three first terms in equation (4) should be included. This is further supported by the short response time discussed in section 6.

On field lines with $L$ value $< \sim 6\ R_E$, the field again becomes symmetric. *Stenbaek-Nielsen and Otto* [1997] suggested that due to Ampere's law, a gradient of $B_y$ along the $x$ axis implies a current along the $z$ axis (see Figure 7), forming an interhemispheric current. This mechanism is shown in Figure 7 for the case of IMF $B_y$ positive. The direction of the currents in Figure 7 is consistent with those shown in Figures 6a' and 6b'.

However, we do not agree with the terminology used in *Stenbaek-Nielsen and Otto* [1997], and we argue that the system can be properly understood only when the dynamics are included. We argue that the currents proposed by *Stenbaek-Nielsen and Otto* [1997] are not interhemispheric currents. They are not interhemispheric in the sense that they cross the neutral sheet (or equatorial plane) or that the pressure balance exists between the foot points in each hemisphere.

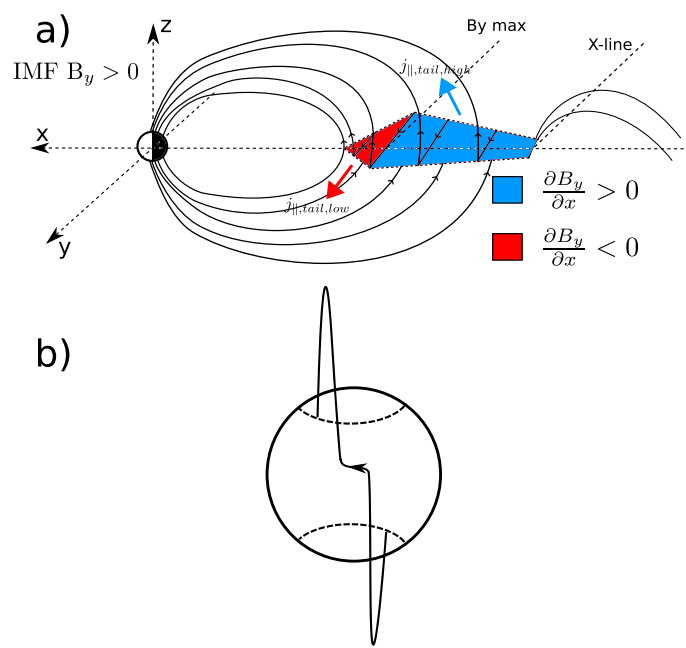

**Figure 7.** (a) Illustration of induced $B_y > 0$ in the closed magnetosphere and how the pileup region results in a gradient in $B_y$, which would according to the authors result in an interhemispheric current (remake of *Stenbaek-Nielsen and Otto* [1997, Figure 4]). (b) View from the magnetotail toward the Sun.

Instead, we consider this as two pairs of currents systems. The opposite-directed stresses from the magnetospheric source region (Figures 6a' and 6b' red and blue arrows) are exerted on each ionosphere. However, the force balance does not exist between the two ionospheres but between the near-Earth magnetosphere and each ionosphere separately. The magnitude of the currents in these two current systems is determined by the forces and energy flux which are asymmetrically shared between the magnetospheric source plasma and the two ionospheres. By arguments given above there should be no net Birkeland current flowing between the two hemispheres. Instead, a pair of flux tubes carrying balanced Birkeland currents closes by transverse currents in each ionosphere as well as in the source region near the equator (Figure 6). In Figure 6 the magnitude of the currents in the two systems has different magnitude due to the pressure surrounding Earth ($\nabla P_0$) opposing the stress in one hemisphere and is parallel in the other. We note that any additional ionospheric asymmetries, such as differences in conductivity between the hemispheres, could enhance or mitigate the asymmetry of the current magnitudes.

We suggest that the term interhemispheric currents should be reserved to explain situations where one ionosphere is active (source) and is driving a current to opposite ionosphere. In such a situation the force balance exists between the two foot points, and the current does, in fact, cross the equatorial plane.

## 5. Observation of Asymmetric Birkeland Currents

The Active Magnetosphere and Planetary Electrodynamics Response Experiment (AMPERE) provides global continuous sampling of the magnetic field perturbations [*Anderson et al.*, 2000; *Waters et al.*, 2001; *Korth et al.*, 2004; *Anderson et al.*, 2008]. We constructed statistical maps sorted by IMF $B_y > 0$. We use the stability criteria defined by *Anderson et al.* [2008], requiring only slowly changing currents. We note that the signatures discussed below are also clearly seen without the stability criteria.

The dayside R1 and R2 currents are modulated by the presence of IMF $B_y$. Figure 8 shows a statistical current density map for the Northern and southern Hemispheres. We discuss the Northern Hemisphere and IMF $B_y > 0$ unless stated otherwise. The magnetosphere imposes a flow dawnward on the dayside (see Figure 2). This, in turn, requires a corresponding Pedersen current. This closure current is transmitted by Alfvèn waves, and the associated Birkeland currents serve to transmit the energy and momentum to the ionosphere. Upward current corresponds to downward electrons flowing into the ionosphere, and we expect auroral signatures. Opposite behavior in the Southern Hemisphere results in asymmetric aurora. Ionospheric flows are a consequence of momentum transport from the magnetosphere and therefore must be associated with Birkeland current closing in the ionosphere. In the Northern Hemisphere, one observes a "R1" current flowing into (blue) the ionosphere around noon around $\lambda \sim 75°$ and a current flowing out of the ionosphere (red) at higher latitudes. These are signatures of an imposed dawn-directed magnetospheric flow, set up immediately after dayside reconnection by the tension along the field line. The resulting $\vec{J} \times \vec{B}$ force accelerates the plasma dawnward in the Northern Hemisphere (see Figure 2) and duskward in the Southern Hemisphere. This describes the sunward part of the banana convection cell in the northern dawn region (see section 3.3). The dusk region convection cell has the characteristic orange shape.

On the nightside the signatures are seen as a rotation of the current systems between the hemispheres. The relative rotation is about 2 h MLT, represented by the dashed lines in Figure 8. In order to compare the current density between the hemispheres, we must first assess the asymmetry of the conjugate foot points ($\Delta$ MLT). That is, in order to compare the current density at conjugate positions, we must first determine the asymmetry of the foot point (related to the twisting of the field) of the field lines. Studies by *Østgaard et al.* [2004, 2011b] and *Reistad et al.* [2013] suggest $\Delta$MLT ~ 1 h; this MHD model suggests approximately the same; Tsyganenko models suggest only a fraction of that [*Østgaard et al.*, 2007].

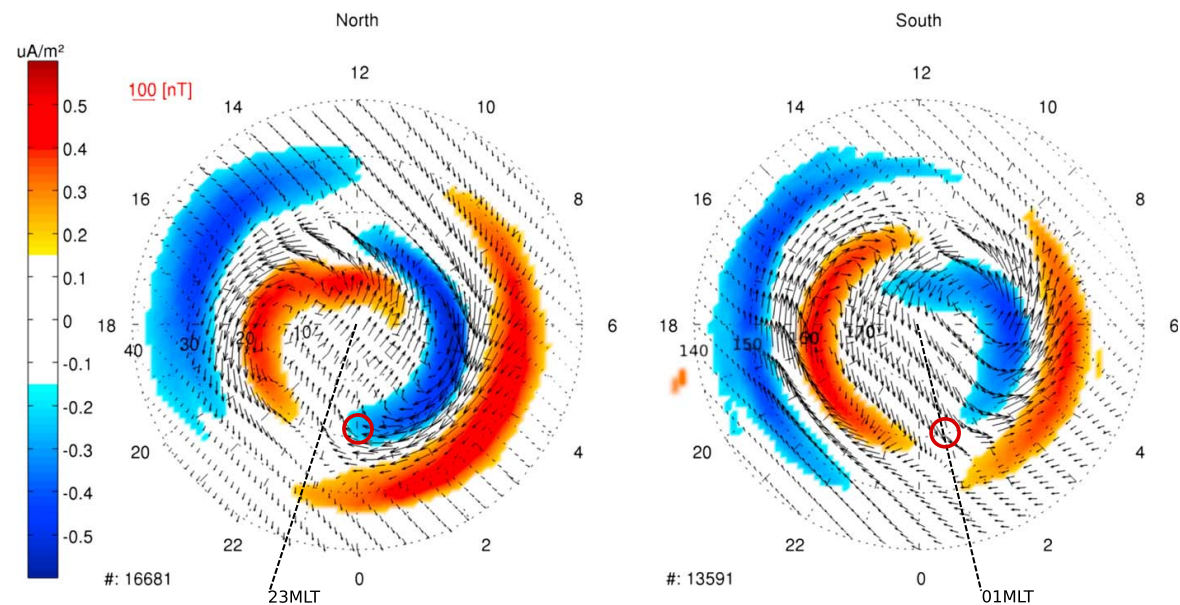

**Figure 8.** Statistical AMPERE maps for $B_z < 0$ and $B_y > 5$ and a stability criterion > 0.45 [see *Anderson et al.*, 2008] in the (left) Northern and (right) Southern Hemispheres. Colors indicate current density (red: away from and blue: into ionosphere), and arrows are the measured magnetic perturbations. The R1 and R2 current systems are rotated about 2 h MLT on the nightside compared to non-$B_y$ conditions as visualized with the dashed lines at ~23 MLT and ~1 MLT. The red circles represent an example of two conjugate locations where the current density can be compared between the hemispheres.

The two red circles in Figure 8 represent two comparable regions under the assumption of $\Delta$MLT = 1 h. With the assumption in mind, we believe that asymmetries in the nightside currents are seen in the AMPERE maps. This suggests asymmetric azimuthal flows, which have been observed with Super Dual Auroral Radar Network [e.g., *Grocott et al.*, 2007].

A study by [*Østgaard et al.*, 2011b] showed the evolution of the asymmetry of the foot points (measured by $\Delta$MLT) during a substorm. The authors found that during the expansion phase of the substorms analyzed the asymmetry disappears. Although, we do not agree with their theoretical arguments, in particular the presence of a net field-aligned current between hemispheres and the role of $\Delta E_{\parallel}$ in the asymmetry of the ionospheric motion. Their observations are consistent with the mechanism we have presented, which is based on magnetic tension and asymmetric azimuthal flows between the two hemispheres.

## 6. Response Time

An important question to understand the dynamics of the system is as follows: how long does it take to produce the induced $B_y$ in the magnetotail?

In our simulation model input the IMF $B_y$ changes from 0 to 10 nT as a simple step function. Both the magnetosphere and ionosphere respond to the change by reconfiguring into a state consistent with the new IMF. The reconfiguration time depends on density and magnetic field strength and is therefore different in the various regions in the magnetosphere. Our estimates presented below are considered as the response time for reaching a new equilibrium.

On the dayside, the twisting of the field lines will evolve as more and more fluxes are removed from northern dusk and added to northern dawn until it saturates with some "efficiency." The closed field lines are affected on the same timescale as the open flux is added to the magnetosphere. Figure 3 shows that after only ~10 min the pressure is distributed asymmetrically; this will in turn force the closed field lines to twist effectively inducing a $B_y$ on timescales less than 15 min. Figure 1b shows the dayside closed field lines at $L$ ~7.5 $R_E$ approximately 20 min after IMF $B_y$ arrived. On the nightside we have presented model runs where we follow the field lines in time from the dayside reconnection and to the magnetotail.

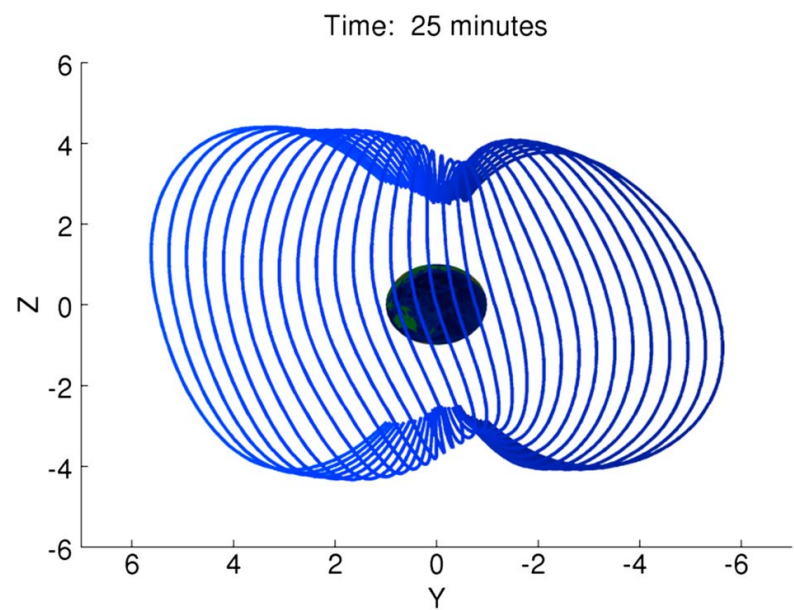

**Figure 9.** Closed field lines on nightside at $L = 11\ R_E$, 25 min after the arrival of IMF $B_y$.

The model suggests that a newly reconnected field line on the dayside undergoes tail reconnection after approximately 45 min. However, the buildup of asymmetric pressure in the lobes is established on much shorter timescales. We argue that the magnetic pressure induced by the asymmetrical loading of flux would induce a $B_y$ on the existing closed field lines. This implies that the asymmetric pressure will induce a $B_y$ on the already present field lines long before the actual field line reconnects in the tail. It is the compressional waves that propagate perpendicular to the magnetic field that affects existing magnetic field lines. The speed of fast-mode waves has been reported to be about 900 km/s [*Wilken et al.*, 1982]. We therefore argue that the time it takes to induce a $B_y$ in the closed magnetosphere is significantly less than the actual convection time from dayside to nightside reconnection (see Figure 2). Figure 9 shows nightside closed field lines at $L = 11\ R_E$ just 25 min after IMF $B_y$ arrived at the dayside magnetopause. The closed field lines are clearly twisted, and the induced $B_y$ has been established on timescales less than the convection time from the dayside magnetopause to nightside reconnection. Observationally, the lag time of the response from the arrival of IMF at the magnetopause has been shown to be about 10 min on the nightside [*Wing et al.*, 2002]. In a study comparing the IMF clock angle and substorm onset location, *Østgaard et al.* [2011a] found highest correlation when time shifting the data <10 min, which is also consistent with what we have described and the model results. The timing of the magnetospheric and convection response has also been thoroughly discussed by *Ruohoniemi et al.* [2002]. A future project will use spacecraft data and modeling to confirm the response timing.

## 7. Summary

In the current work we have addressed how IMF $B_y$ modulates the dynamics and morphology in the magnetosphere and the resulting signatures in the ionosphere.

The process of inducing $B_y$ on nightside closed field lines originates on the dayside. When the IMF possess a $B_y$ component, flux will be added asymmetrically to the magnetosphere via the dayside reconnection process. The governing equations describing the coupling between magnetospheric flow and magnetic tension and pressure are discussed in section 2. The asymmetric distribution of magnetic flux between the hemispheres creates shear flows, acting to restore equilibrium, and will influence existing field lines on both the dayside and nightside, effectively inducing a $B_y$ component. The Maxwell stress exerted on the ionosphere by the tension and pressure of the displaced magnetic flux bundle distorts the ionospheric convection patterns into the characteristic banana and orange cell patterns. We have emphasized the importance of including the dynamical behavior of the system. As the field line convects from the midtail region, it initially experiences asymmetric lobe forces, which twists the field lines. At a later time in the evolution it starts to experience the pressure surrounding Earth which forces the field line to convect either dawnward or duskward, while the lobe pressure is reduced. Eventually, the field line relaxes and tension in field line (induced by the lobe pressure) is released mostly to one of the hemispheres. The result is asymmetric Birkeland currents at conjugate foot points, consistent with the convection patterns. The magnetosphere's response to the impact of IMF $B_y$ has been discussed, and from the modeling perspective we argue that $B_y$ is induced on dayside on timescales of a few minutes and in tens of minutes for the nightside. We presented statistical current density maps from AMPERE. We believe that asymmetric currents at conjugate foot points on the nightside are present in the statistical maps. The underlying assumption is that the rotation of the current systems is smaller than the relative shift of the conjugate foot points, since it is the current density in the conjugate foot points that must be compared. Several studies have reported on fast nightside azimuthal flows associated with IMF $B_y$ [e.g., *Pitkänen et al.*, 2015; *Grocott et al.*, 2007, and references therein].

Below we summarize the impact of IMF $B_y$ on the magnetosphere and ionosphere system:

1. In the presence of IMF $B_y \neq 0$, dayside reconnection results in a asymmetrical distribution of flux between the hemispheres.
2. The asymmetric magnetic pressure in the lobes leads to asymmetric plasma flows and thereby induces $B_y$ on closed field lines on both the dayside and nightside and forces asymmetric displacement of the ionospheric foot points.
3. In the ionosphere, asymmetric azimuthal flows arise, manifested as convection on a banana cell in one hemisphere and an orange cell in the other.
4. $B_y$ is induced independently of nightside reconnection. It occurs on timescales less than 20 min for nightside closed field lines.
5. Asymmetric Birkeland currents (connected to different MLT locations in ionosphere) form as a consequence of the field line convecting from a region of dominating magnetic lobe pressure to an area where the pressure surrounding Earth is dominating. The tension is released into the hemisphere where the tension and Earth pressure act in the same direction.
6. For IMF $B_y > 0$ we expect stronger currents in the northern postmidnight dawn region and in the southern premidnight dusk region. For IMF $B_y < 0$ the signatures are reversed.
7. Rather than interhemispheric currents, we argue that the induced $B_y$ results in pairs of asymmetric Birkeland currents.
8. Signatures of asymmetric currents are seen in AMPERE, primarily as relative rotation of hemispheric patterns.

**Acknowledgments**
We acknowledge the use of NASA/GSFC's Space Physics Data Facility for OMNI data. Simulation results have been provided by the Community Coordinated Modeling Center at Goddard Space Flight Center through their public Runs on Request system (http://ccmc.gsfc.nasa.gov). The CCMC is a multiagency partnership between NASA, AFMC, AFOSR, AFRL, AFWA, NOAA, NSF, and ONR (Paul-Tenfjord-032514-1). We thank the AMPERE team and the AMPERE Science Center for providing the Iridium-derived data products. This study was supported by the Research Council of Norway/CoE under contract 223252/F50.

## References

Anderson, B. J., K. Takahashi, and B. A. Toth (2000), Sensing global Birkeland currents with Iridium engineering magnetometer data, *Geophys. Res. Lett.*, 27(24), 4045–4048, doi:10.1029/2000GL000094.
Anderson, B. J., H. Korth, C. L. Waters, D. L. Green, and P. Stauning (2008), Statistical Birkeland current distributions from magnetic field observations by the Iridium constellation, *Ann. Geophys.*, 26, 671–687, doi:10.5194/angeo-26-671-2008.
Burch, J. L., P. H. Reiff, J. D. Menietti, R. Heelis, W. B. Hanson, S. D. Shawhan, E. G. Shelley, M. Sugiura, D. R. Weimer, and J. D. Winningham (1985), IMF By-dependent plasma flow and Birkeland currents in the dayside magnetosphere: 1. Dynamics explorer observations, *J. Geophys. Res.*, 90, 1577–1593.
Cao, J., A. Duan, M. Dunlop, W. Xinhua, and C. Cai (2014), Dependence of IMF $B_z$ penetration into the neutral sheet on IMF $B_z$ and geomagnetic activity, *J. Geophys. Res.*, 119, 5279–5285, doi:10.1002/2014JA019827.
Cowley, S. (1981), Magnetospheric asymmetries associated with the y-component of the IMF, *Planet. Space Sci.*, 29(1), 79–96, doi:10.1016/0032-0633(81)90141-0.
Cowley, S., and W. Hughes (1983), Observation of an IMF sector effect in the Y-magnetic field component at geostational orbit, *Planet. Space Sci.*, 31(1), 73–90, doi:10.1016/0032-0633(83)90032-6.
Cowley, S., J. Morelli, and M. Lockwood (1991), Dependence of convective flows and particle precipitation in the high-latitude dayside ionosphere on the X and Y components of the interplanetary magnetic field, *J. Geophys. Res.*, 96, 5557–5564, doi:10.1029/90JA02063.
Dungey, J. W. (1961), Interplanetary magnetic field and the auroral zones, *Phys. Rev. Lett.*, 6, 47–48, doi:10.1103/PhysRevLett.6.47.
Escoubet, C. P., A. Pedersen, R. Schmidt, and P. A. Lindqvist (1997), Density in the magnetosphere inferred from ISEE 1 spacecraft potential, *J. Geophys. Res.*, 102(A8), 17,595–17,609, doi:10.1029/97JA00290.
Fairfield, D. H. (1979), On the average configuration of the geomagnetic tail, *J. Geophys. Res.*, 84, 1950–1958, doi:10.1029/JA084iA05p01950.
Friis-Christensen, E., K. Lassen, J. Wilhjelm, J. M. Wilcox, W. Gonzalez, and D. S. Colburn (1972), Critical component of the interplanetary magnetic field responsible for large geomagnetic effects in the polar cap, *J. Geophys. Res.*, 77(19), 3371–3376, doi:10.1029/JA077i019p03371.
Grocott, A., and S. E. Milan (2014), The influence of IMF clock angle timescales on the morphology of ionospheric convection, *J. Geophys. Res. Space Physics*, 119, 5861–5876, doi:10.1002/2014JA020136.
Grocott, A., T. Yeoman, S. Milan, O. Amm, H. Frey, L. Juusola, R. Nakamura, C. Owen, H. Rème, and T. Takada (2007), Multi-scale observations of magnetotail flux transport during IMF-northward non-substorm intervals, *Ann. Geophys.*, 25(2002), 1709–1720, doi:10.5194/angeo-25-1709-2007.
Haaland, S. E., G. Paschmann, M. Förster, J. M. Quinn, R. B. Torbert, C. E. McIlwain, H. Vaith, P. A. Puhl-Quinn, and C. A. Kletzing (2007), High-latitude plasma convection from Cluster EDI measurements: Method and IMF-dependence, *Ann. Geophys.*, 25, 239–253, doi:10.5194/angeo-25-239-2007.
Hau, L., and G. Erickson (1995), Penetration of the interplanetary magnetic field $B_y$ into Earth's plasma sheet, *J. Geophys. Res.*, 100(2), 745–751.
Heppner, J. P. (1972), Polar-cap electric field distributions related to the interplanetary magnetic field direction, *J. Geophys. Res.*, 77(25), 4877–4887, doi:10.1029/JA077i025p04877.
Heppner, J. P., and N. Maynard (1987), Empirical high latitude electric field models, *J. Geophys. Res.*, 92, 4467–4489.
Iijima, T. (2000), Field-aligned currents in geospace: Substance and significance, in *Magnetospheric Current Systems*, edited by S. Ohtani et al., pp. 107–129, AGU, Washington, D. C., doi:10.1029/GM118p0107.
Kabin, K., R. Rankin, R. Marchand, T. I. Gombosi, C. R. Clauer, A. J. Ridley, V. O. Papitashvili, and D. L. DeZeeuw (2003), Dynamic response of Earth's magnetosphere to $B_y$ reversals, *J. Geophys. Res.*, 108(A3), 1132, doi:10.1029/2002JA009480.
Kaymaz, Z., G. L. Siscoe, J. G. Luhmann, R. P. Lepping, and C. T. Russell (1994), Interplanetary magnetic field control of magnetotail magnetic field geometry: IMP 8 observations, *J. Geophys. Res.*, 99, 11,113–11,126, doi:10.1029/94JA00300.
Khurana, K. K., R. J. Walker, and T. Ogino (1996), Magnetospheric convection in the presence of interplanetary magnetic field $B_y$: A conceptual model and simulations, *J. Geophys. Res.*, 101, 4907–4916.

Korth, H., B. J. Anderson, M. J. Wiltberger, J. G. Lyon, and P. C. Anderson (2004), Intercomparison of ionospheric electrodynamics from the Iridium constellation with global MHD simulations, *J. Geophys. Res.*, 109, A07307, doi:10.1029/2004JA010428.

Kozlovsky, A. (2003), IMF $B_y$ effects in the magnetospheric convection on closed magnetic field lines, *Geophys. Res. Lett.*, 30(24), 2261, doi:10.1029/2003GL018457.

Kullen, A., and L. O. Blomberg (1996), The influence of IMF on the mapping between the Earth's magnetotail and its ionosphere, *Geophys. Res. Lett.*, 23(18), 2561–2564.

Laakso, H., R. Pfaff, and P. Janhunen (2002), Polar observations of electron density distribution in the Earth's magnetosphere: 1. Statistical results, *Ann. Geophys.*, 20(11), 1711–1724, doi:10.5194/angeo-20-1711-2002.

Liou, K., and P. T. Newell (2010), On the azimuthal location of auroral breakup: Hemispheric asymmetry, *Geophys. Res. Lett.*, 37, L23103, doi:10.1029/2010GL045537.

Liou, K., P. T. Newell, D. G. Sibeck, C.-I. Meng, M. Brittnacher, and G. Parks (2001), Observation of IMF and seasonal effects in the location of auroral substorm onset, *J. Geophys. Res.*, 106, 5799–5810.

Lui, A. T. Y. (1984), Characteristics of the cross-tail current in the Earth's magnetotail, in *Magnetospheric Currents*, edited by T. A. Potemra, pp. 158–170, AGU, Washington, D. C., doi:10.1029/GM028p0158.

Lyatskaya, S., W. Lyatsky, and G. Khazanov (2014), Distinguishing high surf from volcanic long-period earthquakes, *Geophys. Res. Lett.*, 41, 799–804, doi:10.1002/2013GL058954.

Lyon, J. G., J. A. Fedder, and C. M. Mobarry (2004), The Lyon-Fedder-Mobarry (LFM) global MHD magnetospheric simulation code, *J. Atmos. Sol. Terr. Phys.*, 66(15–16), 1333–1350, doi:10.1016/j.jastp.2004.03.020.

Mansurov, S. M. (1969), New evidence of the relationship between magnetic field in space and on the Earth [English Translation], *Geomagn. Aeron.*, 9, 768–773.

Merkin, V. G., and J. G. Lyon (2010), Effects of the low-latitude ionospheric boundary condition on the global magnetosphere, *J. Geophys. Res.*, 115, A10202, doi:10.1029/2010JA015461.

Milan, S. E. (2015), Sun et lumière: Solar wind-magnetosphere coupling as deduced from ionospheric flows and polar auroras, in *Magnetospheric Plasma Physics: The Impact of Jim Dungey's Research, Astrophys. Space Sci. Proc.*, vol. 41, edited by D. Southwood, S. W. H. Cowley, and S. Mitton, pp. 33–64, Springer, Switzerland, doi:10.1007/978-3-319-18359-6.

Mozer, F. S., C. F. Bogott, M. C. Kelley, S. Schutz, and W. D. Gonzalez (1974), High-latitude electric fields and the three-dimensional interaction between the interplanetary and terrestrial magnetic fields, *J. Geophys. Res.*, 79(1), 56–63, doi:10.1029/JA079i001p00056.

Østgaard, N., and K. M. Laundal (2012), Auroral asymmetries in the conjugate hemispheres and interhemispheric currents, in *Auroral Phenomenology and Magnetospheric Processes: Earth And Other Planets, Geophys. Monogr. Ser.*, edited by A. Keiling et al., pp. 99–111, AGU, Washington, D. C., doi:10.1029/2011GM001190.

Østgaard, N., S. B. Mende, H. U. Frey, T. J. Immel, L. A. Frank, J. B. Sigwarth, and T. J. Stubbs (2004), Interplanetary magnetic field control of the location of substorm onset and auroral features in the conjugate hemispheres, *J. Geophys. Res.*, 109, A07204, doi:10.1029/2003JA010370.

Østgaard, N., S. B. Mende, H. U. Frey, J. B. Sigwarth, A. Åsnes, and J. M. Weygand (2007), Auroral conjugacy studies based on global imaging, *J. Atmos. Sol. Terr. Phys.*, 69, 249–255, doi:10.1016/j.jastp.2006.05.026.

Østgaard, N., K. M. Laundal, L. Juusola, A. Åsnes, S. E. Håland, and J. M. Weygand (2011a), Interhemispherical asymmetry of substorm onset locations and the interplanetary magnetic field, *Geophys. Res. Lett.*, 38, L08104, doi:10.1029/2011GL046767.

Østgaard, N., B. K. Humberset, and K. M. Laundal (2011b), Evolution of auroral asymmetries in the conjugate hemispheres during two substorms, *Geophys. Res. Lett.*, 38, L03101, doi:10.1029/2010GL046057.

Ouellette, J. E., B. N. Rogers, M. Wiltberger, and J. G. Lyon (2010), Magnetic reconnection at the dayside magnetopause in global Lyon-Fedder-Mobarry simulations, *J. Geophys. Res.*, 115, A08222, doi:10.1029/2009JA014886.

Park, K. S., T. Ogino, and R. J. Walker (2006), On the importance of antiparallel reconnection when the dipole tilt and IMF $B_y$ are nonzero, *J. Geophys. Res.*, 111, A05202, doi:10.1029/2004JA010972.

Parker, E. N. (1996), The alternative paradigm for magnetospheric physics, *J. Geophys. Res.*, 101, 10,587–10,625, doi:10.1029/95JA02866.

Parker, E. N. (2007), *Conversations on Electric and Magnetic Fields in the Cosmos*, Princeton Univ. Press, Princeton, N. J.

Petrukovich, A. A. (2011), Origins of plasma sheet $B_y$, *J. Geophys. Res.*, 116, A07217, doi:10.1029/2010JA016386.

Petrukovich, A. A., W. Baumjohann, R. Nakamura, A. Runov, and A. Balogh (2005), Cluster vision of the magnetotail current sheet on a macroscale, *J. Geophys. Res.*, 110, A06204, doi:10.1029/2004JA010825.

Pitkänen, T., M. Hamrin, P. Norqvist, T. Karlsson, H. Nilsson, A. Kullen, S. M. Imber, and S. E. Milan (2015), Azimuthal velocity shear within an earthward fast flow further evidence for magnetotail untwisting?, *Ann. Geophys.*, 33, 245–255, doi:10.5194/angeo-33-245-2015.

Provan, G., M. Lester, S. Mende, and S. E. Milan (2009), Statistical study of high-latitude plasma flow during magnetospheric substorms, *Ann. Geophys.*, 22, 3607–3624, doi:10.5194/angeo-22-3607-2004.

Reistad, J. P., N. Østgaard, K. M. Laundal, and K. Oksavik (2013), On the non-conjugacy of nightside aurora and their generator mechanisms, *J. Geophys. Res. Space Physics*, 118, 3394–3406, doi:10.1002/jgra.50300.

Reistad, J. P., N. Østgaard, K. M. Laundal, S. Haaland, P. Tenfjord, K. Snekvik, and K. Oksavik (2014), Intensity asymmetries in the dusk sector of the poleward auroral oval due to IMF $B_x$, *J. Geophys. Res. Space Physics*, 119, 9497–9507, doi:10.1002/2014JA020216.

Ridley, A. J., T. I. Gombosi, I. V. Sokolov, G. Tóth, and D. T. Welling (2010), Numerical considerations in simulating the global magnetosphere, *Ann. Geophys.*, 28(8), 1589–1614, doi:10.5194/angeo-28-1589-2010.

Ruohoniemi, J., S. Shepherd, and R. Greenwald (2002), The response of the high-latitude ionosphere to IMF variations, *J. Atmos. Sol. Terr. Phys.*, 64, 159–171, doi:10.1016/S1364-6826(01)00081-5.

Russel, C. (1972), The configuration of the magnetosphere, in *Critical Problems of Magnetospheric Physics*, edited by E. R. Dyer, pp. 1–16, IUSTP Secretariat, Washington, D. C.

Song, P., and V. M. Vasyliünas (2010), Aspects of global magnetospheric processes, *Chin. J. Space Sci.*, 30(4), 289–311.

Song, P., and V. M. Vasyliunas (2011), How is the ionosphere driven by the magnetosphere?, in *General Assembly and Scientific Symposium, 2011 XXXth URSI*, pp. 1–4, IEEE, doi:10.1109/URSIGASS.2011.6050873, 13–20 Aug.

Stenbaek-Nielsen, H. C., and A. Otto (1997), Conjugate auroras and the interplanetary magnetic field, *J. Geophys. Res.*, 102(A2), 2223–2232, doi:10.1029/96JA03563.

Strangeway, R. J. (2012), The relationship between magnetospheric processes and auroral field-aligned current morphology, in *Auroral Phenomenology and Magnetospheric Processes: Earth And Other Planets*, edited by A. Keiling et al., pp. 355–364, AGU, Washington, D. C., doi:10.1029/2012GM001211.

Svalgaard, L. (1968), Sector structure of the interplanetary magnetic field and daily variation of the geomagnetic field at high latitudes, *Geophys. Pap. R-16*, Danish Meteorol. Inst.Copenhagen, Denmark.

Tsyganenko, N. A. (1998), Modeling of twisted/warped magnetospheric configurations using the general deformation method, *J. Geophys. Res.*, 103, 23,551–23,563, doi:10.1029/98JA02292.

Tu, J., P. Song, and V. M. Vasyliunas (2014), Inductive-dynamic magnetosphere-ionosphere coupling via MHD waves, *J. Geophys. Res. Space Physics*, *119*, 530–547, doi:10.1002/2013JA018982.

Vasyliunas, V. M. (2012), The physical basis of ionospheric electrodynamics, *Ann. Geophys.*, *30*(2), 357–369, doi:10.5194/angeo-30-357-2012.

Walker, R. J., R. L. Richard, T. Ogino, and M. Ashour-Abdalla (1999), The response of the magnetotail to changes in the IMF orientation: The magnetotail's long memory, *Phys. Chem. Earth Part C*, *24*(1–3), 221–227, doi:10.1016/S1464-1917(98)00032-4.

Walsh, A. P., et al. (2014), Dawn-dusk asymmetries in the coupled solar wind-magnetosphere-ionosphere system: A review, *Ann. Geophys.*, *32*, 705–737, doi:10.5194/angeo-32-705-2014.

Waters, C. L., B. J. Anderson, and K. Liou (2001), Estimation of global field aligned currents using the Iridium System magnetometer data, *Geophys. Res. Lett.*, *28*(11), 2165–2168, doi:10.1029/2000GL012725.

Wilken, B., C. K. Goertz, D. N. Baker, P. R. Higbie, and T. A. Fritz (1982), The SSC on July 29, 1977 and its propagation within the magnetosphere, *J. Geophys. Res.*, *87*(A8), 5901–5910, doi:10.1029/JA087iA08p05901.

Wing, S., P. T. Newell, D. G. Sibeck, and K. B. Baker (1995), A large statistical study of the entry of interplanetary magnetic field Y-component into the magnetosphere, *Geophys. Res. Lett.*, *22*(16), 2083–2086, doi:10.1029/95GL02261.

Wing, S., D. G. Sibeck, M. Wiltberger, and H. Singer (2002), Geosynchronous magnetic field temporal response to solar wind and IMF variations, *J. Geophys. Res.*, *107*(A8), 1222, doi:10.1029/2001JA009156.